\documentclass[10pt,journal,compsoc]{IEEEtran}

\usepackage{setspace}
\usepackage{placeins}
\usepackage{afterpage}
\usepackage[utf8]{inputenc}
\usepackage[cmex10]{amsmath}
\usepackage{graphicx}
\usepackage{amssymb}
\usepackage{amsthm}
\usepackage{cite}
\usepackage{color}
\usepackage{caption}
\usepackage{amsmath}
\usepackage{tabu}
\usepackage{array}
\usepackage{booktabs}
\usepackage{amssymb}
\usepackage{threeparttable}
\usepackage[linesnumbered,ruled]{algorithm2e}
\usepackage{float}
\usepackage{stfloats}
\usepackage{lipsum}
\usepackage[english]{babel}

\usepackage[T1]{fontenc}
\usepackage{mwe}    
\usepackage{subfig}

\let\oldnl\nl
\newcommand{\nonl}{\renewcommand{\nl}{\let\nl\oldnl}}

\theoremstyle{definition}

\begin{document}

\title{On the Importance of Location Privacy for Users of Location Based Applications}

\author{Sina Shaham, Saba Rafieian, Ming Ding, Mahyar Shirvanimoghaddam, and Zihuai Lin
	\thanks{S. Shaham, M. Shirvanimoghaddam and Z. Lin are with the Department
		of Engineering, The University of Sydney, Sydney, NSW, 2006 Australia (e-mail: sina.shaham, Mahyar.Shirvanimoghaddam, zihuai.lin\}@sydney.edu.au).}
	\thanks{S. Rafieian is with The Ryerson University, Toronto, ON,  M5B2K3 Canada (email: saba.rafieian@ryerson.ca)}
	\thanks{M. Ding is with Data61, Sydney, NSW, 1435 Australia (email: ming.ding@data61.csiro.au)}

	}

\IEEEtitleabstractindextext{
	\begin{abstract}
    		
    Do people care about their location privacy while using location-based service apps? This paper aims to answer this question and several other hypotheses through a survey, and review the privacy preservation techniques. \textbf{Our results indicate that privacy is indeed an influential factor in the selection of location-based apps by users}. 
	\end{abstract}
	\begin{IEEEkeywords}
		Location Privacy, Technology Acceptance Model
\end{IEEEkeywords}}
\maketitle

\IEEEraisesectionheading{\section{Introduction}\label{Introduction}}

\IEEEPARstart{W}{ith} the tremendous growth of smart devices and mobile internet in the last decade, mobile apps have become an integral part of our everyday life and a lucrative business for service providers. Same as how users highly depend on the services mobile apps offer, service providers are thirsty for more data from users to provide a better experience and gain more customers. One crucial source of information is the users' location data. Apps that use the location of users to provide a service are usually referred to as Location-Based Service (LBS) apps. Through these applications, users submit their locations to LBS providers and in return, benefit from the services they offer. An example of an LBS app is Google Maps, with over 2 billion monthly users in 2018. It is not a surprise to know that the annual market for LBSs is expected to reach 68.85 billion US dollars by 2023, according to ‘Research and Markets’ report~\cite{c1}. Fig.~1 categorizes the services LBS apps provide based on their most widely used applications.

We have demonstrated the widely accepted system model for LBSs in Fig.~2. Users make queries on the LBS app, providing their location, and requesting for a service. The telecommunications network transmits queries to the LBS provider replying with the requested services. Luckily, the telecommunication infrastructure is considered to be trusted as governments have enforced regulations to protect the privacy of users. However, such protection and laws are not easily applied or even exist for LBS providers. If an LBS provider aims to exploit users' location data, they may accumulate the requested location queries over time and apply data mining techniques to infer sensitive information about users. For example, LBS providers can learn about the home address, shopping habits, and the users' daily routines. Such threats have sparked the importance of location privacy.

Duckham et al.~\cite{c2} define location privacy as the ability to prevent other parties from learning one’s current or past locations. In simple words, location privacy refers to having control over how our location data is being used. Recent misconduct in the US election and Facebook scandal has exposed the importance of privacy and in particular, location privacy more than any time before in history.

\begin{figure}[t]
\centering
\includegraphics[scale=.48]{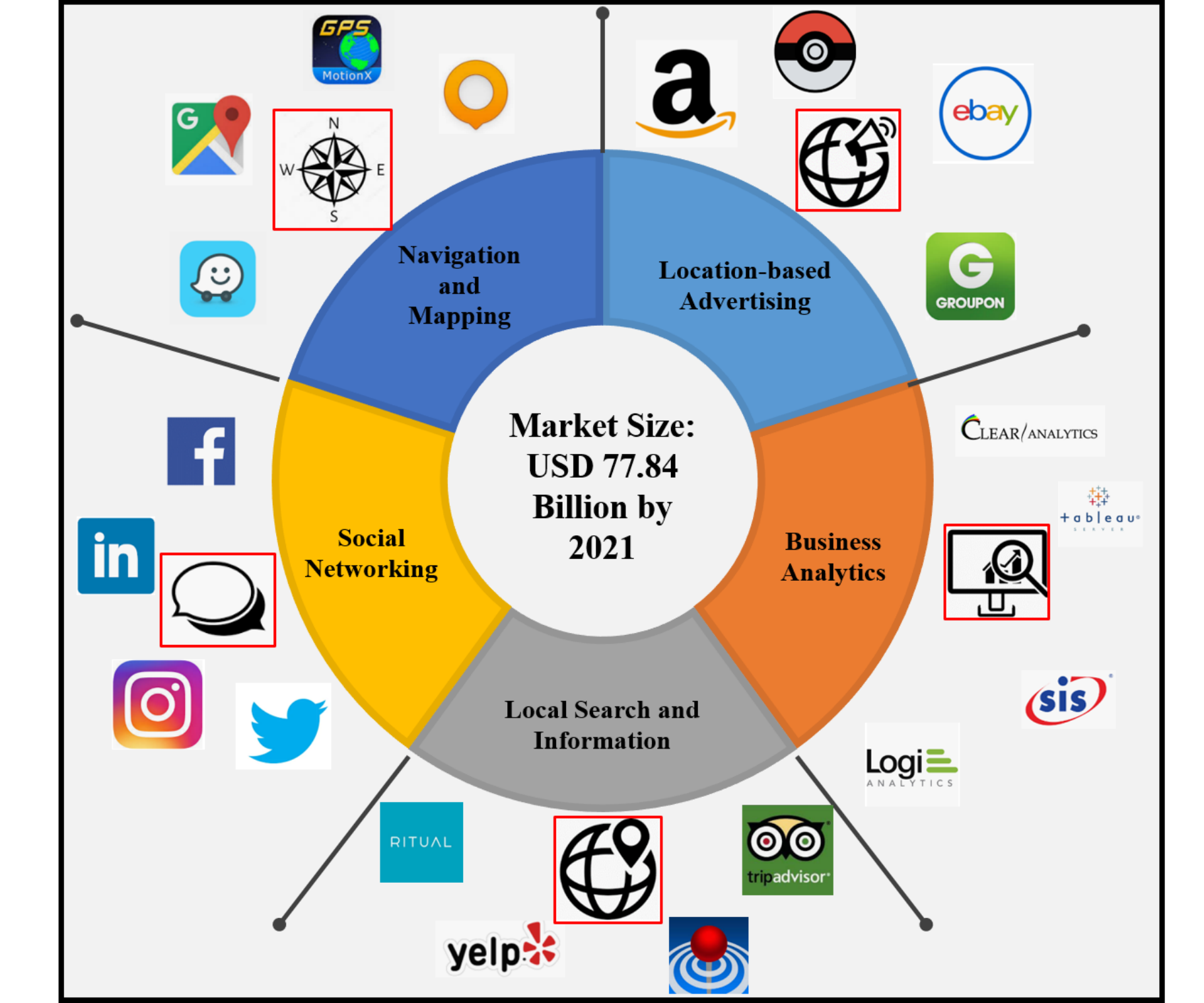}
\hspace{1em}
\centering
\caption{Classification of LBS apps based on their application.}\label{f1}
\end{figure}

Although there is a consensus in the research community on the significance of location privacy and privacy preservation in LBS apps, it is crucial to understand the users’ perspective. Do people care about their location privacy while using LBS apps? Does location privacy have any impact on their decision to use an LBS app? In this work, we take a step further to understand if location privacy actually affects the choice of apps that users download and use. Our primary hypothesis is that the perceived privacy of LBS apps from users’ perspectives directly affects their intention to use the apps. We designed a survey to test our hypothesis. Our study is not limited to the implications of the perceived privacy of LBS apps. We provide a comprehensive analysis of factors affecting the acceptance of LBS applications examining seven hypotheses to understand user insights. We also review the state of the art techniques that can be used to improve the location privacy of users in addition to presenting potential future research directions and challenges.

\begin{figure}[t]
\centering
\includegraphics[scale=.7]{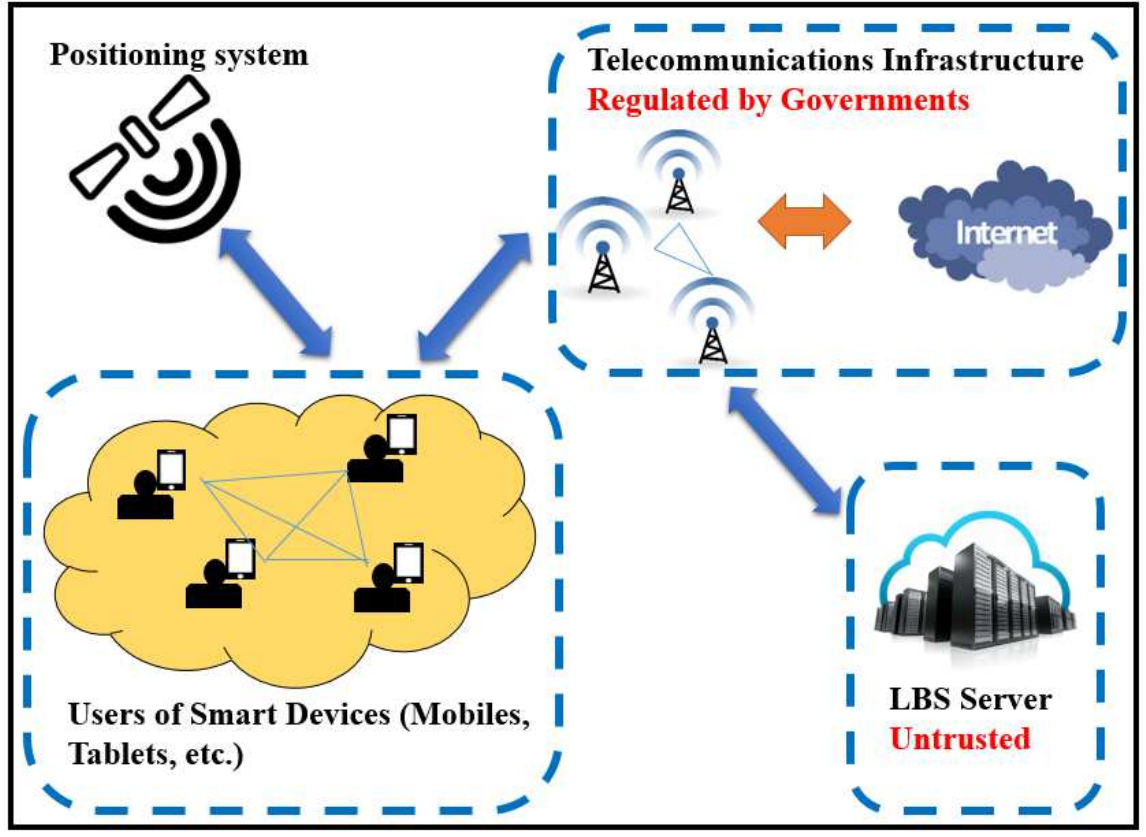}
\hspace{1em}
\centering
\caption{System model for users of LBSs.}\label{f2}
\end{figure}

\section{Do People Care About Location Privacy?}\label{Do People Care About Location Privacy?} 

To find out the importance of location privacy in the eyes of the users and whether it influences their decision to choose an LBS application, we designed a questionnaire based on the Technology Acceptance Model (TAM). TAM was initially proposed by~\cite{c3} and widely applied in different companies and enterprises to understand the user acceptance of new technologies. Just a few examples of using TAM are:

\begin{itemize}
    \item Examining physicians’ decisions on the acceptance of telemedicine technology at public hospitals in Hong Kong~\cite{c4}.
    \item Examining user acceptance of hotel reservation websites for both hotel-owned and third parties~\cite{c5}.
    \item Examining external factors influencing user acceptance of personalized online shopping~\cite{c6}.
\end{itemize}

Fig.~3 demonstrates our proposed model to understand the user acceptance of LBS apps and the importance of location privacy. In the remaining of this section, we start by explaining how we have conducted our experiments and the methodology used to analyze our hypotheses. Then, we describe each element or so-called construct of the model, including our findings, insights, and discussions.

\begin{figure*}[t]
\centering
\includegraphics[scale=1.2]{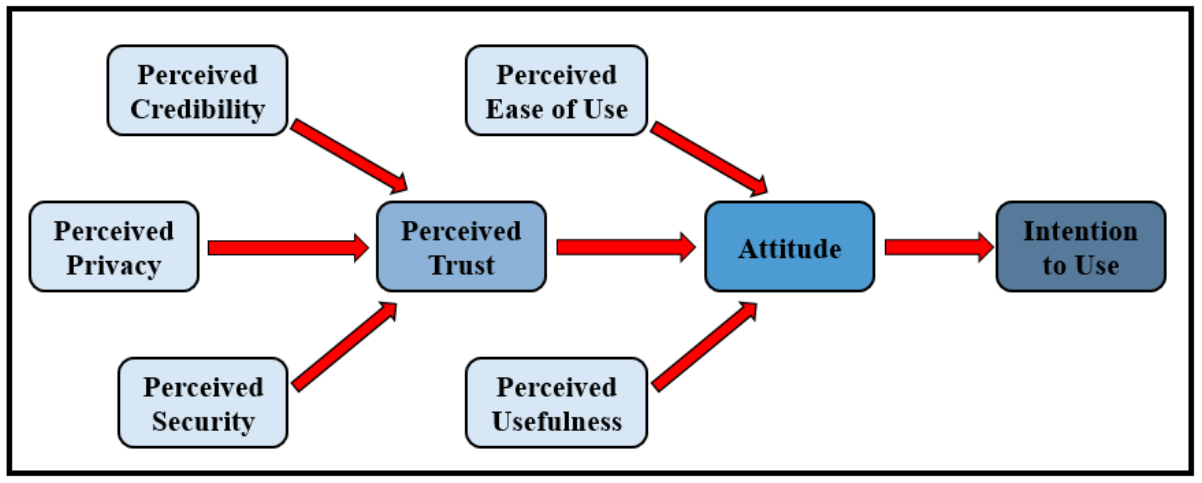}
\hspace{1em}
\centering
\caption{Our proposed TAM for analyzing the user perspective towards their intention to use of LBS applications.}\label{f3}
\end{figure*}

\subsection{Methodology}\label{Methodology} 
We conducted our survey among 98 people based on random sampling and carefully analyzed the answers to understand the attitude of users towards LBS apps. The participants of the survey consisted of 58 male and 40 female respondents. To understand the cause and effect relationships and evaluate our hypotheses, we analyzed the data using the Partial Least Square Path Model (PLS-PM) with the aid of SmartPLS software. We verified 
\begin{itemize}
    \item \textbf{RELIABILITY} of our data using Cronbach's alpha and composite reliability~\cite{c7}. Cronbach's alpha indicates how well the questions used for each construct of our model measures the construct. All our Cronbach's Alpha values are close to or higher than 0.7 for each construct, which is the widely accepted threshold to consider questions as reliable. We further investigated the reliability of our data using composite reliability, which indicates the internal consistency of the questions for each construct. The acceptable benchmark of the minimum 0.7 achieved for this measure in all constructs.
    
    \item \textbf{VALIDITY} of our measurements using convergent validity and discriminant validity~\cite{c8}. Convergent validity is an indicator that proves the relations between questions that were supposed to be related theoretically. To have convergent validity, there are two requirements; First, the factor loadings should be higher than 0.7 as it is the case for our questionnaire. Second, the average variance extracted should be higher than 0.5, which is also valid based on our collected data. Next, we considered discriminant validity to prove that measured constructs which should not be related theoretically, are actually not related. For this to be true, the squared root of variance shared between constructs and their items needs to be higher than the correlations between the construct and any other construct, which is also true for our designed questionnaire.
    
\end{itemize}

\subsection{Results \& Discussions}
Our model in Fig.~3 shows the cause and effect relationships that we hypothesized to exist between the constructs of our model. Each arrow indicates our belief in the significance of one construct on another. For instance, we hypothesize that the attitude of users has a significant influence on the intention to use of the LBS apps; therefore, an arrow is originated from the 'attitude' construct to 'intention to use' construct. To have a better understanding of the relationships between constructs, Fig.~4 provides the scatter plot for the considered cause and effect relationships. The background color of plots describes if the relation between the considered two variables is statistically insignificant, significant with low influence, or significant with high influence. In the remaining of this section, we define each construct of the model and explain our findings corresponding to that construct, and finally present a summary of our most important findings.

	\begin{figure*}[!ht]
		\subfloat[\label{dis1}]{%
			\includegraphics[scale=1.12]{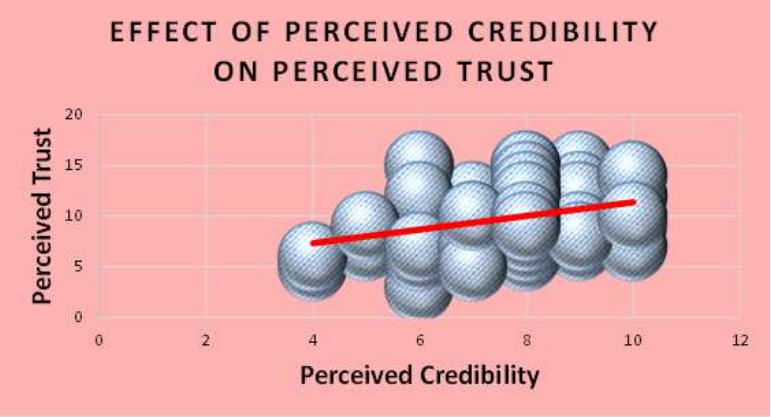}
		}
		\hfill
		\subfloat[\label{dis2}]{%
			\includegraphics[scale=1.12]{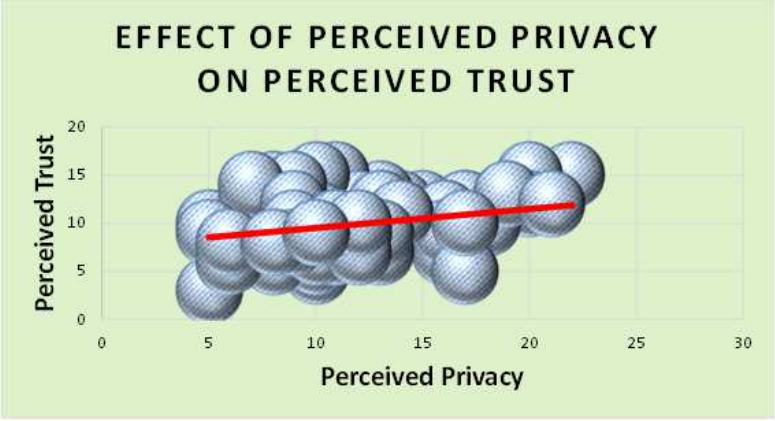}
		}
		\hfill
		\subfloat[\label{dis3}]{%
		\includegraphics[scale=1.12]{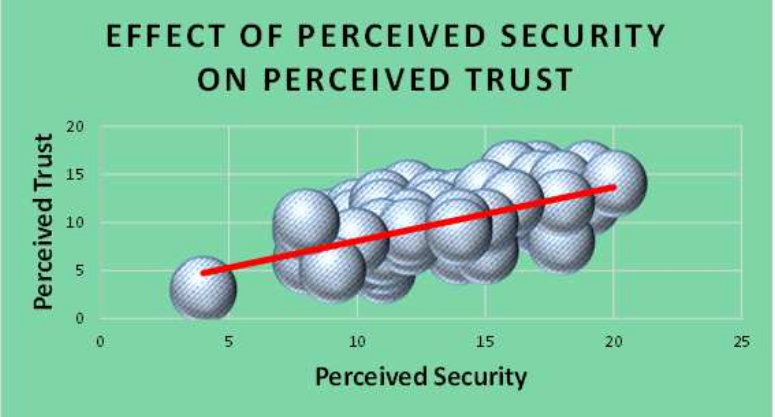}
		}
		\hfill
		\subfloat[ \label{query4}]{%
		\includegraphics[scale=1.12]{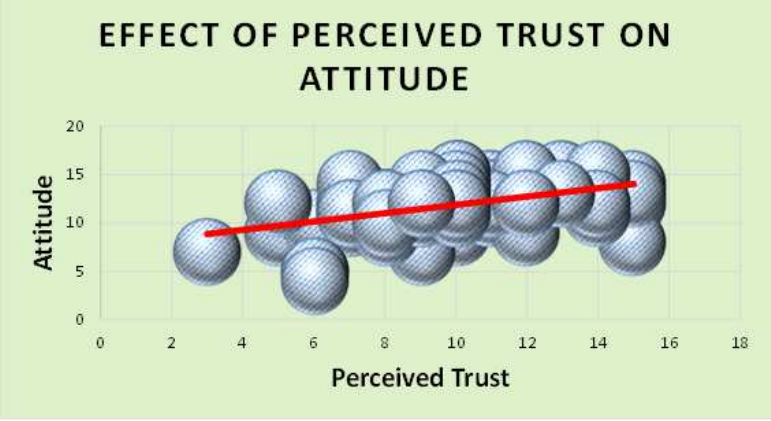}
		}
		\hfill
		\subfloat[ \label{query5}.]{%
		\includegraphics[scale=1.12]{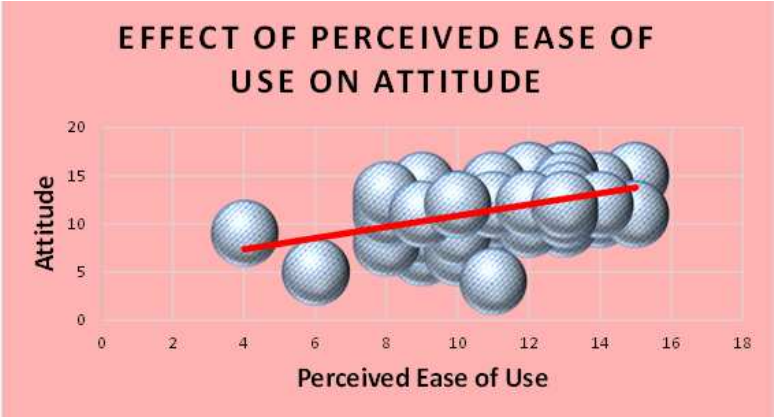}
		}
		\hfill
		\subfloat[ \label{error6}]{%
		\includegraphics[scale=1.12]{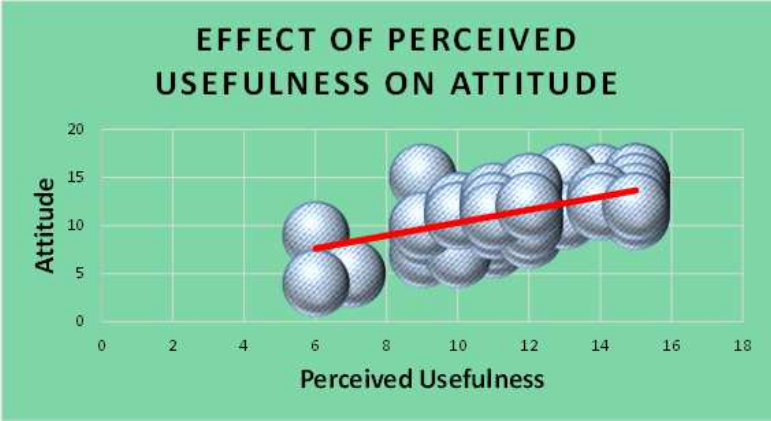}
		}
		\hfill
		\subfloat[\label{len7}]{%
		\includegraphics[scale=1.12]{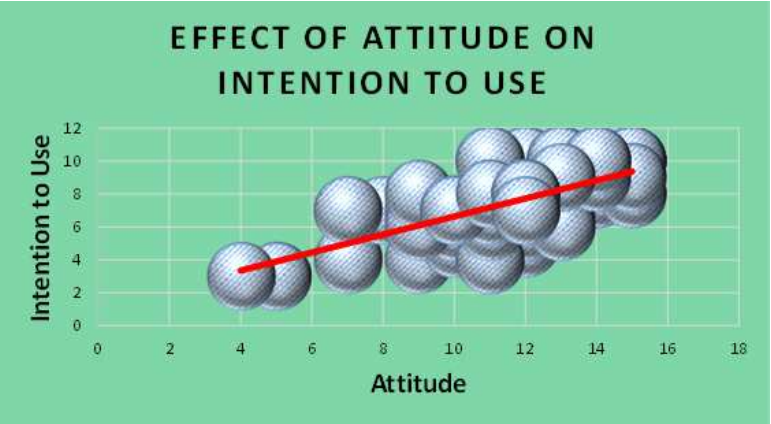}
		}
		\hfill
		\centering
	    \captionsetup[subfigure]{labelformat=empty}
		\subfloat[\label{time8}]{%
		\includegraphics[scale=1.3]{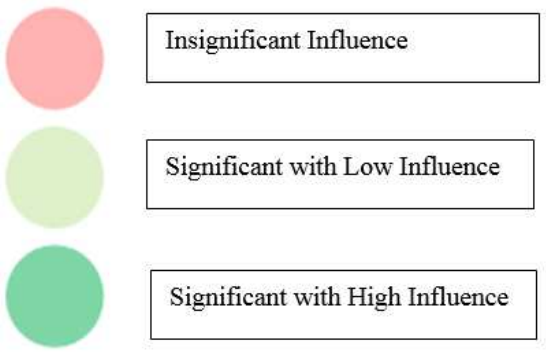}
		}
		\caption{Visualization of Cause and effect relationships existing between the constructs. The background color of diagrams represents how significant the influence of one variable is on another.}
		\label{fig:dummy1}
	\end{figure*}

\textbf{Perceived Credibility} indicates users' perception of how efficiently and reliably their given tasks and requested services from LBS providers can be handled. Our initial hypothesis was that perceived credibility should have a significant effect on the perceived trust of the LBS apps. Our findings indicate that it is not the case. The impact of perceived credibility is not high enough on the perceived trust of the users to be considered as significant (Fig.~4a). This is good news for new competitors entering the LBS market as their lack of reputation cannot substantially hinder the success of their application in the market.

\textbf{Perceived Privacy} indicates how in control the users think they are while giving their information to LBS providers regarding the way their data is being acquired and used. According to the results of our survey, perceived privacy has a significant effect on the perceived trust of the users (Fig.~4b). The impact of perceived privacy is found to be less than perceived security but still significant. Our finding points out that people are concerned about how service providers are using their data. Therefore, it is crucial to devise approaches for preserving the location privacy of people in LBS apps.

\textbf{Perceived Security} is the belief of users on the preservation of integrity, confidentiality, and non-recognition of their personal information given to service providers. As expected, our findings show that the perceived security of the LBS applications is the most critical factor influencing the perceived trust of the customers (Fig.~4c). This means that the users of LBS applications care more for security compared to other factors in the model. A major reason for this observation could be the lack of awareness of the threats that a breach of privacy can cause. 

\textbf{Perceived Trust} states how secure the users feel while depending on the LBS providers. According to our results, perceived trust has a significant influence on the attitude of the users towards LBS apps (Fig.~4d). Therefore, it counts for a substantial component on the decision that the users make about using the LBS apps. As explained, the perceived trust of users is significantly influenced by perceived security and perceived privacy of the LBS apps. 

\textbf{Perceived Ease of Use and Perceived Usefulness} constructs are crucial constructs used in our model, which are proposed as part of the original TAM. Perceived ease of use represents the degree that the users think using LBS applications are free of efforts, and perceived usefulness refers to the level that the users believe the use of LBS apps would help them to enhance their job and everyday performance. According to our survey, perceived usefulness of the LBS apps has the highest impact on the attitude of users (Fig.~4f).  However, perceived ease of use of the LBS apps is not found to have a significant effect on the attitude of users towards using LBS apps (Fig.~4e). 

\textbf{Attitude (A) \& Intention to Use (IU)} are the ultimate constructs of the model indicating whether users are going to use an LBS app or not. As two major factors of the original TAM model, we hypothesized that the attitude of users significantly influences their intention to use of LBS applications, which expectedly turns out to be the case (Fig.~4g). In our model, we anticipated that three constructs affect the attitude of users: perceived ease of use, perceived usefulness, and perceived trust, among which perceived usefulness and trust of the LBS apps found to be the most influential factors.\\

In summary, the key insights that our model provides are: 

\begin{itemize}
    \item Perceived trust of users has a significant impact on the attitude of users towards using LBS apps. \\
    \item Perceived privacy of users towards using LBS apps has a significant impact on their intention to use.\\
    \item Perceived usefulness of LBS apps is found to be the most critical factor influencing the intention to use.\\
    \item Perceived credibility and perceived ease of use of LBS apps are not found to be significant factors affecting the intention to use of the LBS apps.
\end{itemize}

\section{Techniques To Improve Location Privacy}\label{system model}

As our experiment suggests, location privacy is not just technical jargon, but it has a real influence on the attitude of individuals. In this section, we review some of the techniques that can improve users' location privacy. Most of these approaches aim at achieving a privacy metric called k-anonymity, which indicates that the location of users is not identifiable from at least k-1 other users~\cite{c9}. As shown in Fig.~2, the LBS provider can act as an adversary and compromise the location privacy of users. It can apply data mining techniques to infer sensitive information about users and take advantage of such information by selling them to other parties or using them for other purposes, such as advertisement. Therefore, the following techniques are developed by researchers to protect users against such threats. Note that all the following methods correspond to the application layer of smart devices.

\textbf{Generation of dummy locations} to hide the location of users from LBS providers is proposed to preserve the privacy of users~\cite{c10}. The idea is to request services for k-1 fake locations as well as the real location of users. Such locations are referred to as dummy locations. Dummy locations help to confuse the LBS provider in identifying the actual location of users as they are also queried with the same user identifier. The significance of the dummy generation technique is that there is no need for a trusted service provider to anonymize users' location data, as the existence of such trusted service providers can become a problem for the privacy of users itself.

\textbf{Pseudonym assignment} is a technique to hide the identity of users with the aid of a trusted service provider. Identifier of a user can be the name, IP address, or any other property that can be uniquely related to users. In this technique, the trusted service provider collects the location information of users and assigns a fake identifier or so-called pseudonym, and then, transfers it to the LBS provider. Therefore, the trusted service provider prevents the LBS provider, which is considered to be untrusted from learning users’ identity. As finding a truly trusted service provider can become a problem itself, some in the literature suggest creating the pseudonyms by the users in the network.

\textbf{Mixed zones} technique focuses on specifying areas in which users are not identifiable~\cite{c11}. Once a user enters a mixed zone, its identifier changes to a predefined pseudonym specified by a trusted service provider. Therefore, once a service is requested from a user in the mixed zone, the LBS provider would not be able to identify who has asked for the service. Applying the mixed zones idea has attracted much attention from vehicular communications. In some scenarios, pseudonym change technique and mixed zones are combined to provide further privacy for the users of smart devices.

\textbf{Location cloaking} is another popular technique to preserve the privacy of users while benefiting from numerous advantages of LBS apps~\cite{c12}. This technique also requires a trusted third party to preserve the location privacy of users. The trusted server receives queries from the users and generates a cloaking box around them, including k-1 other locations, and then queries them altogether from the LBS provider. Consequently, the technique aims to achieve k-anonymity for users by requesting the service for a group of locations based on the same identifier. 

\textbf{Cryptography} approaches, such as the one proposed in~\cite{c13}, are becoming more and more popular among the researchers. This technique is implemented as part of the Private Information Retrieval process, in which users retrieve information from the server without it learning what information has been requested. Most approaches~\cite{c14,c15} in this area consider the technique in a theoretical setting. Hence, they call for more research to reduce the computational complexity and enable cryptography to protect the location privacy of users.

\section{Challenges \& Future Directions}\label{Experiments}

In broader view of the preserving location privacy of the users, techniques can be categorized into two groups: (I) approaches that only involve the users and LBS provider or so-called two-tier spatial transformation, and (II) approaches that require a trusted anonymizer as well as the users and LBS provider, or so-called three-tier spatial transformation. The first category involves techniques such as cryptography and a dummy-based approach. This category provides an additional layer of privacy as no other parties are involved that could possibly comprise the privacy of users. However, a principal assumption behind dummy-based algorithms in two-year spatial transformation techniques is that adversaries do not hold any background knowledge that can help them to identify the users. For instance, if the adversary knows about the probability of queries being made from cells, it is shown that it can apply probabilistic or correlation attacks to identify the users. This does not hold for the cryptography technique. If future research is able to reduce the complexity of the cryptography technique, it can be implemented in practical scenarios and has the potential to guarantee the privacy of users in telecommunication networks.

The techniques based on three-tier spatial transformation, such as location cloaking and mixed zones, can provide higher levels of privacy compared to dummy-based algorithms, even if adversaries have background knowledge about users. However, the cost of such an enhanced performance is higher runtime overhead as users need to update their location data with a third-party before they can benefit from LBSs provided by the LBS server. Moreover, many of the algorithms proposed in this category result in delays and high computational complexity as well as a lower quality of service. Additionally, introducing an entirely trusted third-party server can become a point of contention itself.

\section{Conclusion}\label{conclusion}

In this article, we investigated the importance of location privacy from the user perspective. According to our findings, privacy has a significant impact on the decision of users to use an LBS app. Therefore, it is crucial to develop robust techniques to improve the location privacy of users.

\bibliographystyle{IEEEtran}
\bibliography{Magazine}

\begin{thebibliography}{10}
\providecommand{\url}[1]{#1}
\csname url@samestyle\endcsname
\providecommand{\newblock}{\relax}
\providecommand{\bibinfo}[2]{#2}
\providecommand{\BIBentrySTDinterwordspacing}{\spaceskip=0pt\relax}
\providecommand{\BIBentryALTinterwordstretchfactor}{4}
\providecommand{\BIBentryALTinterwordspacing}{\spaceskip=\fontdimen2\font plus
\BIBentryALTinterwordstretchfactor\fontdimen3\font minus
  \fontdimen4\font\relax}
\providecommand{\BIBforeignlanguage}[2]{{%
\expandafter\ifx\csname l@#1\endcsname\relax
\typeout{** WARNING: IEEEtran.bst: No hyphenation pattern has been}%
\typeout{** loaded for the language `#1'. Using the pattern for}%
\typeout{** the default language instead.}%
\else
\language=\csname l@#1\endcsname
\fi
#2}}
\providecommand{\BIBdecl}{\relax}
\BIBdecl

\bibitem{c1}
\BIBentryALTinterwordspacing
``Location based service market 2018 global industry trends, statistics, size,
  share, growth factors, emerging technologies, regional analysis, competitive
  landscape forecast to 2023.'' [Online]. Available:
  \url{https://www.thefreelibrary.com/Location Based Service Market 2018 Global
  Industry Trends}
\BIBentrySTDinterwordspacing

\bibitem{c2}
M.~Duckham and L.~Kulik, ``Location privacy and location-aware computing,'' in
  \emph{Dynamic and Mobile GIS}.\hskip 1em plus 0.5em minus 0.4em\relax CRC
  press, 2006, pp. 63--80.

\bibitem{c3}
F.~D. Davis, ``Perceived usefulness, perceived ease of use, and user acceptance
  of information technology,'' \emph{MIS quarterly}, pp. 319--340, 1989.

\bibitem{c4}
P.~J. Hu, P.~Y. Chau, O.~R.~L. Sheng, and K.~Y. Tam, ``Examining the technology
  acceptance model using physician acceptance of telemedicine technology,''
  \emph{Journal of management information systems}, vol.~16, no.~2, pp.
  91--112, 1999.

\bibitem{c5}
C.~Morosan and M.~Jeong, ``Users’ perceptions of two types of hotel
  reservation web sites,'' \emph{International Journal of Hospitality
  Management}, vol.~27, no.~2, pp. 284--292, 2008.

\bibitem{c6}
D.~Baier and E.~St{\"u}ber, ``Acceptance of recommendations to buy in online
  retailing,'' \emph{Journal of Retailing and Consumer Services}, vol.~17,
  no.~3, pp. 173--180, 2010.

\bibitem{c7}
M.~Sarstedt, C.~M. Ringle, and J.~F. Hair, ``Pls-sem: Looking back and moving
  forward,'' 2014.

\bibitem{c8}
C.~Fornell and D.~F. Larcker, ``Evaluating structural equation models with
  unobservable variables and measurement error,'' \emph{Journal of marketing
  research}, vol.~18, no.~1, pp. 39--50, 1981.

\bibitem{c9}
L.~Sweeney, ``k-anonymity: A model for protecting privacy,''
  \emph{International Journal of Uncertainty, Fuzziness and Knowledge-Based
  Systems}, vol.~10, no.~05, pp. 557--570, 2002.

\bibitem{c10}
H.~Kido, Y.~Yanagisawa, and T.~Satoh, ``An anonymous communication technique
  using dummies for location-based services,'' in \emph{ICPS'05. Proceedings.
  International Conference on Pervasive Services, 2005.}\hskip 1em plus 0.5em
  minus 0.4em\relax IEEE, 2005, pp. 88--97.

\bibitem{c11}
A.~R. Beresford and F.~Stajano, ``Location privacy in pervasive computing,''
  \emph{IEEE Pervasive computing}, no.~1, pp. 46--55, 2003.

\bibitem{c12}
M.~Gruteser and D.~Grunwald, ``Anonymous usage of location-based services
  through spatial and temporal cloaking,'' in \emph{Proceedings of the 1st
  international conference on Mobile systems, applications and services}.\hskip
  1em plus 0.5em minus 0.4em\relax ACM, 2003, pp. 31--42.

\bibitem{c13}
G.~Ghinita, P.~Kalnis, A.~Khoshgozaran, C.~Shahabi, and K.-L. Tan, ``Private
  queries in location based services: anonymizers are not necessary,'' in
  \emph{Proceedings of the 2008 ACM SIGMOD international conference on
  Management of data}.\hskip 1em plus 0.5em minus 0.4em\relax ACM, 2008, pp.
  121--132.

\bibitem{c14}
B.~Chor, O.~Goldreich, E.~Kushilevitz, and M.~Sudan, ``Private information
  retrieval,'' in \emph{Proceedings of IEEE 36th Annual Foundations of Computer
  Science}.\hskip 1em plus 0.5em minus 0.4em\relax IEEE, 1995, pp. 41--50.

\bibitem{c15}
E.~Kushilevitz and R.~Ostrovsky, ``Replication is not needed: Single database,
  computationally-private information retrieval,'' in \emph{Proceedings 38th
  Annual Symposium on Foundations of Computer Science}.\hskip 1em plus 0.5em
  minus 0.4em\relax IEEE, 1997, pp. 364--373.

\end{thebibliography}

\end{document}